\documentclass[english,aps,prc, twocolumn,nofootinbib,0superscriptaddress]{revtex4-1}
\usepackage{epsfig}
\usepackage{graphicx}
\usepackage{amsmath}
\usepackage{lipsum}
\usepackage{color}
\usepackage[hyperindex=true,colorlinks=false,linkcolor=blue]{hyperref}


\def\orcid#1{\kern .08em\href{https://orcid.org/#1}{\includegraphics[keepaspectratio,width=0.7em]{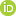}}}

\usepackage{babel}

\begin{document}


\title{Sensitivity of quasi-free reactions on details of the bound-state overlap functions }

\author{C. A. Bertulani \!\!\orcid{0000-0002-4065-6237}}
\affiliation{Department of Physics and Astronomy, Texas A\&M University-Commerce, Commerce, TX 75429-3011, USA}
\affiliation{Institut f\"ur Kernphysik, Technische Universit\"at Darmstadt, D-64289 Darmstadt, Germany}
\author{A. Idini \!\!\orcid{0000-0001-7624-8975} }
\affiliation{Division of Mathematical Physics, Department of Physics, LTH, Lund University, PO Box 118, S-22100 Lund, Sweden}
\author{C. Barbieri \!\!\orcid{0000-0001-8658-6927}}
\affiliation{Dipartimento di Fisica, Universit`a degli Studi di Milano, Via Celoria 16, 20133 Milano, Italy}
\affiliation{INFN, Sezione di Milano, Via Celoria 16, 20133 Milano, Italy}
\affiliation{Department of Physics, University of Surrey, Guildford GU2 7XH, UK}

\begin{abstract}
It is often stated that heavy-ion nucleon knockout reactions are mostly sensitive to the tails of the bound-state wavefunctions. In contrast, (p,2p) and (p,pn) reactions are known  to access information on the full overlap functions within the nucleus. We analyze the  oxygen isotopic chain and explore the differences between single-particle wave functions generated with potential models, used in the experimental analysis of knockout reactions, and \emph{ab initio} computations from self-consistent Green's function theory.  Contrary to the common belief, we find that not only the tail of the overlap functions, but also their internal part are assessed in both reaction mechanisms, which are crucial to yield accurately determined spectroscopic information. 
\end{abstract}

\maketitle

{\bf Introduction.}
High energy ($\gtrsim 100$ MeV/nucleon) neutron and proton removal (knockout) reactions with, e.g., $^9$Be and $^{12}$C targets are one of the most successful tools to investigate the single-particle structure of the many-body wavefunction of nuclei far from the stability.  A large number of experiments yielded an enormous amount of knowledge collected on magicity, shell-evolution, two-- and three--body halo configurations, spectroscopy of deep lying states, etc. The magnitude of the knockout cross sections, as well as the momentum distribution of the fragments, have been the main source of information since the very beginning of this experimental campaign~\cite{ANNE199019,RIISAGER1992365,Orr:92}. Theories have been developed for a credible description of the experimental data~\cite{Bertulani:92,PhysRevC.53.2007,PhysRevC.54.3043,Bertulani:04,Bertulani:06}. 

It is widely considered that knockout reactions are peripheral and probe the tail of the nucleon removal wavefunction, due to absorption at low impact parameters (see, e.g., Refs. \cite{PhysRevLett.77.1016,PhysRevC.100.054607,doi:10.1063/1.4909557,AUMANN2021103847}). The removal wavefunction is given by the overlap integral $I(r) = \left< \Psi^{A-1}_i|\psi(r)|\Psi^{A}_{g.s.}\right>$, where $\left|\Psi^{A}_{g.s.}\right>$ and $\left|\Psi^{A-1}_i\right>$ respectively denote the (many-body) wavefunctions for the projectile and the residual fragment in its $i$-th excited state~\cite{Aumann:00,BoffiBook:96}. The operator $\psi(r)$ removes a nucleon at position~$r$. The tail of $I(r)$ is proportional to the Whittaker function, 
\begin{equation}
I(r) =\left< \Psi^{A-1}_i|\psi(r)|\Psi^{A}_{g.s.}\right> \; \xrightarrow[\; r \rightarrow \infty \;]  \; ~ C {1\over r} W_{-\eta,l+1/2} (2\kappa r) \; , 
\label{eq:overlap}
\end{equation}
where $\kappa=\sqrt{2\mu E_B}/\hbar$ is the wavenumber, $\mu$ the reduced mass between the outgoing nucleon and the $(A-1)$ residual, $E_B$ the removed nucleon separation energy, $\eta=\mu Z_N Z_{(A-1)} e^2/\hbar \kappa$ is the Coulomb parameter, with $Z_A$ and $Z_N$ the target and projectile charges, and $l$ the angular momentum of the removed nucleon. 

Eikonal models for knockout reactions \cite{Bertulani:92,PhysRevC.53.2007,PhysRevC.54.3043,Bertulani:04,Bertulani:06,PhysRevLett.77.1016,PhysRevC.100.054607,doi:10.1063/1.4909557,AUMANN2021103847,Hansen:03} imply that the total knockout cross section is proportional to the integral of the square $I^2(r)$ and, as long as the reaction is truly peripheral, to the squared asymptotic normalization coefficient (ANC): $C^2$. In this case the ANC is the only messenger carrying information about the complex many-body wavefunctions $\left|\Psi^{A}_{g.s.}\right>$ and $\left|\Psi^{A-1}_k\right>$ entering Eq.~\eqref{eq:overlap}. 
\emph{Ab initio} methods compute the shape of overlap functions microscopically, even where these are not well represented by mean-field orbits. 
Moreover, they can handle the large model spaces necessary to resolve the \emph{full} quenching of spectroscopic factors due to correlations~\cite{Barbieri2009prl}. 
In contrast, phenomenological assumptions on radial shapes cannot be avoided even for long-used approaches such as the shell model~\cite{Caurier2005rmp,GadeBrown2008SkX}. 
In practice, most applications in the literature still assume that the tail of $I(r)$ does not differ from an independent-particle approximation (IPA) wavefunction, for example a Woods-Saxon plus spin-orbit tuned to the corresponding separation energy.
When compared to the experimental data of nucleon knockout reactions, the square of $C$ can be extracted and compared to predictions of many-body models (e.g., shell model calculations). This procedure is used to determine the spectroscopic factors $S$ according to~\cite{1990JETPL..51..282M,Hansen:03},
\begin{equation}
C_{\rm exp}^2 = S \cdot C_{\rm IPA}^2 \; ,
\label{eq:anc}
\end{equation}
where $C_{\rm IPA}$ is computed assuming that its IPA wavefunction is properly normalised to unity%
\footnote{Spectroscopic factors are often labelled as $C^2\tilde{S}$ in shell-model and reaction theory. The $\tilde{S}$ represents the quenching of strength due to inter-nucleon correlations while $C^2$ are Clebsh-Gordan coefficients that account for partial occupation of orbits. Here, we follow the convention from the {\sl ab initio} community using $S\equiv C^2\tilde{S}$---to avoid confusions between ANCs and Clebsh-Gordans.}.
Nuclear correlations have the effect of quenching the spectroscopic factor,
$S\equiv \int I^2(r)\,d^3r$, and so the experimental value of $C_{\rm exp}$ is  smaller than its IPA. The many-body ANC, $C_{\rm MB}$, computed through Eq.~\eqref{eq:overlap} should be compared directly to $C_{\rm exp}$.  The primary goal of nucleon knockout experiments with heavy ion targets is to extract information on the spectroscopic factors and the~$C_{\rm MB}$.

\begin{table*}[t]
\caption{Separation energies, $E_B$, root mean square radii of the overlap wavefunction, $\left<r^2\right>^{1/2}$, asymptotic normalization coefficients (ANC), $(p,pN)$ quasi-free cross sections, $\sigma_{qf}$  and nucleon knockout cross sections, $\sigma_{ko}$, with $^9$Be targets, for 350 MeV/nucleon oxygen projectiles. WS denotes wavefunctions calculated with a potential model (Woods-Saxon) and GF denotes many-body \emph{ab initio} overlap functions from self-consistent Green's function method. For a few cases we generated two different WS orbits, with the second choice constrained to reproduce the same radii as GF.
 Different final states are distinguished by their separation energy, $E_B$. The first columns indicates the target isotope and the mean-field WS orbit that could be tentatively associated to the transferred nucleon. 
 $S_{GF}$ are theoretical spectroscopic factors predicted by SCGF, all other results employ overlap functions normalised to unity. }
\begin{tabular}{|l|c|c|c|c|c|c|c|c|c|c|}
\hline\hline
  Nucleus     &  $E_B$ &  $\left<r^2\right>^{1/2}_{WS}$ &$\left<r^2\right>^{1/2}_{GF}$&C$_{WS}$&C$_{GF}$&$\sigma^{WS}_{qf}$&$\sigma^{GF}_{qf}$&$\sigma_{ko}^{WS}$&$\sigma_{ko}^{GF}$&$S_{GF}$\\
(state) & [MeV] & [fm] &[fm]&[fm$^{-1/2}$]&[fm$^{-1/2}$]&[mb]&[mb]&[mb]&[mb]&\\
\tableline
$^{14}$O ($\pi$1p$_{3/2}$) & 8.877 &  2.836 &2.961&6.665&7.060&20.72&21.28&26.28&28.15&0.548\\
$^{14}$O ($\pi$1p$_{1/2}$) & 6.181 &  2.991 &3.160&4.872&5.401&21.08&16.89&28.61&31.33&0.760\\
$^{14}$O ($\nu$1p$_{3/2}$) & 21.33 &  2.513 &2.722&11.39&14.64&30.55&32.80&21.13&23.92&0.773\\
$^{16}$O ($\pi$s$_{1/2}$) & 15.89 &  2.295 &2.233&13.06&13.81&7.870& 7.696&16.97&15.81&0.074\\
$^{16}$O ($\pi$1p$_{3/2}$) & 17.43 &  2.612 &2.832&15.29&18.27&17.41&18.58&19.83&22.70&0.805\\
$^{16}$O ($\pi$1p$_{1/2}$) & 10.65 &  2.816 &3.077&8.624&10.70&9.094&9.913&22.54&26.29&0.794\\
& &  3.077 &&11.22&&9.625&&25.24&&\\
$^{16}$O ($\nu$1p$_{3/2}$) & 20.71 &  2.580 &2.807&11.96&13.88&27.88&30.26&18.81&21.66&0.801\\
$^{16}$O ($\nu$1p$_{1/2}$) & 13.83 &  2.767 &3.032&6.684&7.578&14.64&16.47&21.20&24.89&0.790\\
$^{22}$O ($\pi$1p$_{3/2}$) & 29.26 &  2.554 &2.884&43.74&63.52&14.37&17.08&13.07&14.50&0.274\\
& &  2.884 &&75.87&&15.47&&15.72&&\\
$^{22}$O ($\pi$1p$_{3/2}$) & 25.67 &  2.606 &2.820&35.00&54.07&13.30&14.20&12.93&15.10&0.443\\
&  &  2.820 &&49.22&&15.13&&14.66&&\\
$^{22}$O ($\pi$1p$_{1/2}$) & 23.58 &  2.634 &2.916&30.49&51.49&6.607&7.253&13.27&16.21&0.731\\
$^{22}$O ($\nu$1d$_{5/2}$) & 6.670 &  3.328 &3.533&4.519&4.685&45.30&46.63&21.36&24.28&0.806\\
$^{24}$O ($\pi$1p$_{3/2}$) & 28.57 &  2.609 &2.886&45.76&66.45&12.13&13.29&11.37&14.01&0.675\\
$^{24}$O ($\pi$1p$_{3/2}$) & 31.88 &  2.566 &2.847& 55.88&95.22&11.94&13.11&10.98&13.70&0.042\\
$^{24}$O ($\pi$1p$_{1/2}$) & 25.28 &  2.657 &2.985&37.04&57.21&6.054&6.881&11.81&15.11&0.740\\
$^{24}$O ($\nu$2s$_{1/2}$) & 4.120 &  4.190 &4.479&3.971&4.130&13.94&19.95&31.81&36.45&0.844\\
$^{24}$O ($\nu$1d$_{5/2}$) & 6.961 &  3.436 &3.557&2.056&2.106&40.53&41.95&19.51&21.11&0.832\\
\hline\hline
\end{tabular}
\label{tab1}
\end{table*}

New experiments have been carried out or are planned using $(p,pN)$, with $N = p, n$, reactions in inverse kinematics \cite{Panin:16,Aumann:13,PhysRevLett.120.052501,doi:10.1093/ptep/pty011}. New reaction models have been developed  differing from those appropriate for knockout reactions with heavy targets \cite{Ogata:15,Moro:15}.
The proton probes are more sensitive to the inner parts of the nuclear wavefunction, especially for light nuclear projectiles \cite{Aumann:13}. Since both knockout as well as $(p,pN)$ reactions are notable spectroscopic tools of unstable nuclei, it is imperative to understand to what extent experimental conclusions can be affected by assumptions in modeling $I(r)$.

\begin{figure}[tbp] 
  \centering
  \includegraphics[width=3.in,keepaspectratio]{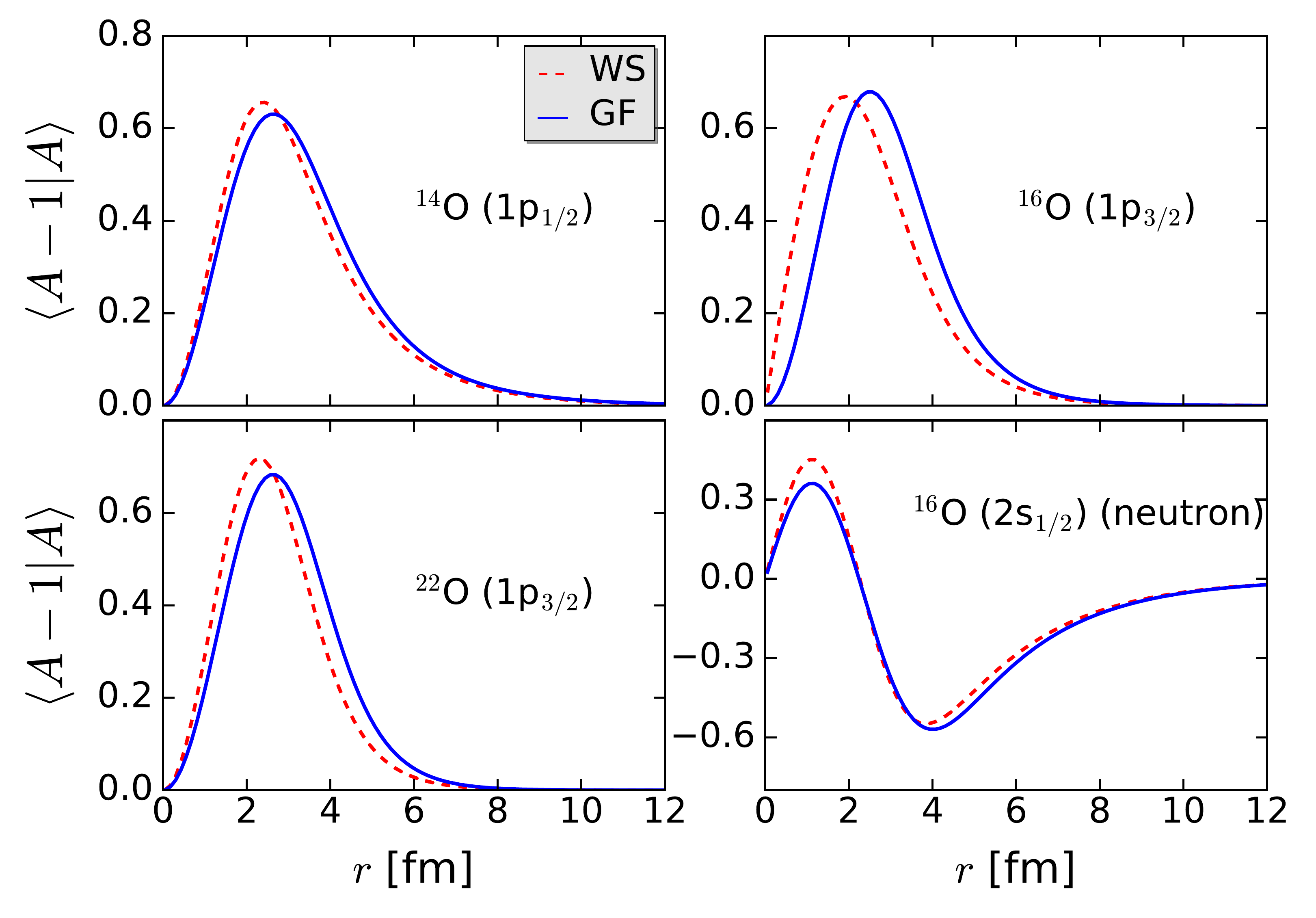}
  \caption{Overlap functions for selected states from Table \ref{tab1}. All cases are for protons except the bottom-right panel for a neutron single-particle state. Solid lines are calculations with the \emph{ab initio} self-consistent Green's function method (GF), while the dashed lines are for Woods-Saxon (WS) potentials reproducing the same separation energies.}
  \label{fig1}
\end{figure}

{\bf Overlaps with \emph{ab initio} methods.}
The \emph{ab initio} overlaps have been calculated from the Hamiltonian
\begin{equation}
 H(A) = T - T_{c.m.}^{[A-1]} + V + W \; ,
\label{eq:H}
\end{equation}
where $T_{c.m.}^{[A-1]}$ is the intrinsic kinetic energy for the recoiling system of mass $A-1$ nucleons, while $V$ and $W$ are the two- and three-body interactions. This formulation is conveniently suited for the calculation of overlap functions and the corresponding nucleon separation energies, $E_h$~\cite{Cipollone:15}.
The three-body term $W$ is reduced to an effective two-body operator as outlined in~\cite{Carbone:13}. We used self-consistent Green's function (SCGF) theory within the third order algebraic diagrammatic construction [ADC(3)] truncation scheme that accounts for all 2p1h, 2h1p intermediate state configurations ~\cite{Schirmer:83,Barbieri:16}. 
The SCGF self-energy was obtained in an harmonic oscillator basis including 14 major shells ($N_{\textrm{max}}=13$) and frequency $\hbar\Omega=20$ MeV. The correct asymptotic tail of our \emph{ab initio} overlap $I^{lj}_{GF}(r)$ 
is ensured by a final Dyson diagonalization in the full (non truncated) momentum space~\cite{Idini:16},
\begin{align}
 \left[E_h - \frac{k^2}{2\mu} \right]\tilde{I}^{lj}_{GF}(k) =&  
 \int\textrm{d}q \, q^2 \, \Sigma^{\star  \,  lj}(k, q; E_h) \, \tilde{I}^{lj}_{GF}(q),
 \label{eq:Schroedinger}
\end{align}
where $\Sigma^{\star}$ is the ADC(3) self-energy, $\mu$ is the reduced mass of the (A-1)-body system plus the ejected nucleon, and $\tilde{I}(k)$ represents the Fourier-Hankel transform of Eq.~\eqref{eq:overlap}.
We perform computations using the NNLO$_\textrm{sat}$ interaction because of its good saturation properties~\cite{Ekstrom:15}. Both radii and binding energies are known to be well reproduced for the oxygen chain nuclei used in this analysis \cite{Lapoux:16}, allowing for a meaningful comparison with reactions from Wood--Saxon--based calculations.

We show the results for $(p,pN)$ quasi-free cross sections using overlap functions obtained with: (a)  SCGF formalism with the chiral NNLO$_\textrm{sat}$ interaction, denoted by $I_{GF}(r)$; (b)  Single particle wavefunctions, $u_{WS}(r)$ generated in a potential model, herewith denoted by $WS$.  The WS radii and diffuseness parameters were taken as $R=1.2A^{1/3}$ fm and $a=0.65$ fm, respectively. A homogeneously charged sphere with radius $R$ was used to generate the Coulomb potential. For case (a) the spectroscopic factors given by $S_{GF} = \int dr I_{GF}^2(r)$ are computed directly from the associated SCGF propagators.  For case (b) the WS model cannot predict the normalization of the overlap functions, hence only empirical spectroscopic factors ($S^{emp}_{WS}$) can be obtained by calculating the quasifree cross sections and comparing to the experimental data.  
Our comparison of the cross section calculations will follow the reaction theory developed in Ref. \cite{Aumann:13} keeping all other input parameters the same, such as separation energies, nuclear densities, etc.

In Table  \ref{tab1} we list 
a series of properties of proton(neutron) knockout reactions
for 350 MeV protons in inverse kinematics, and for oxygen isotopes incident on $^9$Be targets at 350 MeV/nucleon. A selected set of neutron and proton states in oxygen isotopes were chosen. 
For some cases, we included more than one final state for the same nucleus and partial wave removal, corresponding to different excitations of the residual nucleus (hence, different $E_B$). These are computed as distinct correlated (A-1)-nucleon states by SCGF, while we can only assume the same mean-field orbit for all of them if using WS. The shell model explains this fragmentation of the spectrum very well but it falls short of providing microscopic information on the differences between their radial overlaps, similarly to WS. 
The last column in this table lists the spectroscopic factors, as computed from \emph{ab initio} SCGF. To simplify the comparison and focus on the ANC contribution, in this study we keep all GF overlap and WS functions normalised to one, i.e., the cross sections have not been multiplied by the spectroscopic factors $S_{lj}$. In the asymptotic limit (where the nuclear force is vanishingly small), the radial part of the WS wavefunction and GF overlaps can be expressed in terms of Whittaker function  and a corresponding ANC, $C_{lj}$, can be theoretically deduced.  

The r.m.s. radii of the GF wavefunctions are slightly larger than those for the WS wavefunctions.
There seems to be a one-to-one correspondence of this behavior with the quasi-free $(p,pN)$ cross sections which are larger for the GF  wavefunctions. The only exception is the $s_{1/2}$ state in $^{15}$N, fourth row in Table~\ref{tab1}, which does not have dominant single-particle character and cannot be directly associated neither to a $1s_{1/2}$ nor a $2s_{1/2}$ orbit. 
The increase of the cross sections with the r.m.s. radii of the wavefunction is also clearly visible for the additional three other cases (one for $^{16}$O two others for $^{22}$O) where the parameters of the WS potential were adjusted to reproduce the same binding energies and same r.m.s. radii as the GF wavefunctions. The comparison between the cross sections for WS and GF wavefunctions improve, but very noticeable differences remain,  pointing again to the fact that both QFS and knockout reaction mechanisms depend on the details of the wavefunctions.
Similar spectroscopic factors to the one listed in Table (\ref{tab1}) were used in Ref. \cite{PhysRevLett.120.052501} and shown to reproduce the data rather well. It is also worthwhile noticing that in few cases the ANC values are very different between the WS and GF wavefunctions. In essence, {\it it is not necessary, nor expected, that the GF wavefunctions reproduce the same ANC as the WS case because they are constrained by the integral of their internal part, which can vary sensibly due to correlations.}

\begin{figure}[tbp] 
  \centering
  \includegraphics[width=3.in,keepaspectratio]{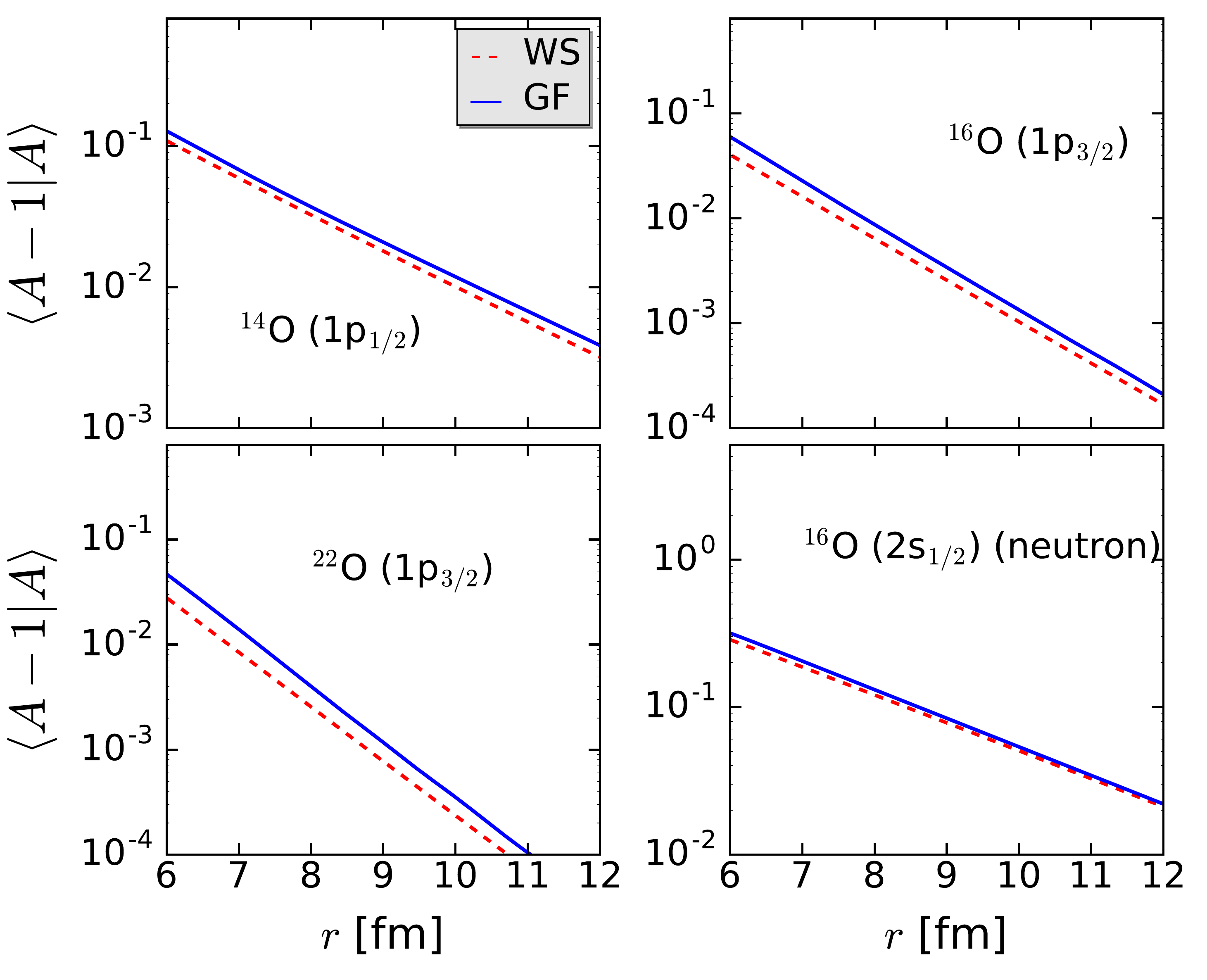}
  \caption{Same as Figure \ref{fig1}, but showing the logarithmic tail of the overlap functions.}
  \label{fig2}
\end{figure}

\begin{figure}[tbp] 
  \centering
  \includegraphics[width=3.in,keepaspectratio]{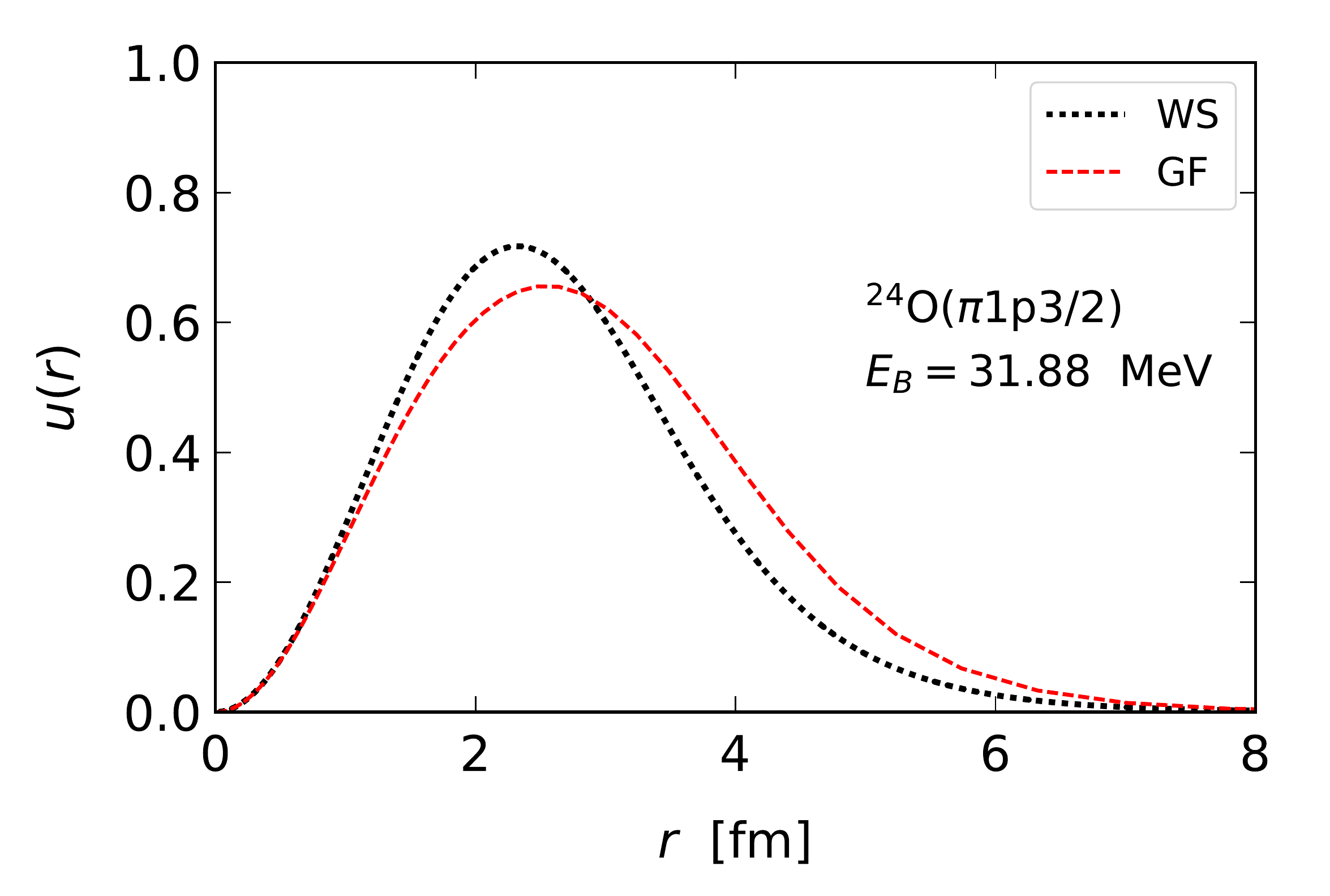}
  \caption{Wood Saxon (WS, red dashed line) wavefunction and Green Function (GF, black dotted line) overlap function for $^{24}$O, 1p3/2, $E_B = 31.88$  MeV.}
  \label{fig-wfOp32-2}
\end{figure}

Earlier \emph{ab initio} wavefunctions obtained from expansions in harmonic oscillator wavefunctions did not reproduce the large distance behavior of the nuclear states unless the expansion runs over a very large number of oscillator shells \cite{PhysRevLett.110.122503,Navratil:06plb,Navratil:06prc}. A simple way to prevent unnecessary large-scale calculations was reported in Refs. \cite{Navratil:06plb,Navratil:06prc} by using a procedure that replaces \emph{ab initio} wavefunctions at their tails by those with appropriate asymptotic behavior such as solutions of a WS model. A fit extending to the internal part of the \emph{ab initio} overlap functions and adequate renormalization yields appropriate values for the ANCs. In fact, it was shown in Refs. \cite{Navratil:06plb,Navratil:06prc} that this procedure leads to an excellent description of cross sections and momentum distributions of proton/neutron knockout reactions with heavy targets based on overlap functions stemming from
the No-Core-Shell-Model (NCSM). Similar issues are now fully resolved for both NCSM~\cite{PhysRevC.87.034326} and for SCGF theories. In our case, the projection of Eq. (4) into momentum space, as discussed in Refs.~\cite{Cipollone:15,PhysRevC.87.034326,Idini:16} always yields the correct asymptotics without the need for ad-hoc corrections.

{\bf Probing deep inside the nucleus.}
An exact reproduction of a Whittaker tail is  irrelevant in for $(p,pN)$ reactions. To show this, in Figure  \ref{fig1} we plot the overlap functions for a few selected states from Table I. All cases are for protons except the bottom-right panel which is for a neutron single-particle state. Solid lines are SCGF calculations, while the dashed lines are for Woods- Saxon (WS) potentials with parameters fitted to match the same separation energies. Evidently, the form of the wavefunctions are not very different, but some difference in the details are noticeable and have an impact on the rms radii and on the quasi-free cross sections, as one can easily read from Table \ref{tab1}. The cross sections can change by as much as 20\%. 

In Figure \ref{fig2} we show the logarithmic tails of the same wavefunctions as in Figure \ref{fig1}. It is clear that our SCGF overlap functions possess very reasonable exponential slopes, as with the WS wavefunctions. Therefore, small differences of the knockout cross sections  in Table \ref{tab1} are due to the authentic modification of the height of the tails due to many-body effects stemming for the interior part of the GF overlap functions. All wavefunctions are normalized to unity.

Substantial differences exist between heavy ion knockout cross sections obtained with single-particle and many-body overlap functions. This can't be ascribed to the asymptotic behavior of the wavefunctions. By simply rescaling the tails of the wavefunction with an ANC or a spectroscopic factor  would lead to a wrong experimental analysis, i.e., {\it just the ANC, or spectroscopic factor, is not enough. The full knowledge of the wavefunction is necessary.} 
 
To clarify the latest point, and show that {\it knockout reactions with heavy ions are also partially sensitive to details of the inner part of the wavefunctions}, consider the probability for one-nucleon stripping in a collision with the core (surviving spectator) having an impact parameter $b$ with the target, while the removed nucleon has an impact parameter $b_n$. The stripping probability is  \cite{Bertulani:04},
\begin{eqnarray}
P_{ko}(b) &=& {\cal S}_c(b)\left< 1- |{\cal S}_n({\bf b}_n)|^2\right> \nonumber \\
&=& {\cal S}_c(b)\int d^3 r |\phi_{nlj}({\bf r})|^2 \left(1- |{\cal S}_n({\bf b}_n\right)|^2)
\label{eq:Pko}
\end{eqnarray}
where $\phi_{nlj}({\bf r})$ denotes the w.f. with
quantum numbers $nlj$ expressed in terms of the relative
core-neutron distance ${\bf r}$. ${\cal S}_c({\cal S}_n)$ is the scattering matrix for the core(nucleon)-target and  $|\phi_{nlj}|^2$ is the probability to find the nucleon
at $\bf r$.
${\bf b}_n\equiv (b_n,\phi_n)$ and the intrinsic coordinate ${\bf r} \equiv (r,\theta,\phi)$ are related by \cite{Bertulani:04}
\begin{eqnarray}
b = \sqrt{r^2 \sin^2\theta + b_n^2 - 2rb_n \sin \theta \cos (\phi-\phi_n)}.
\label{eq:bbn}
\end{eqnarray}

\begin{figure}[tbp] 
  \centering
  \includegraphics[width=3.in,keepaspectratio]{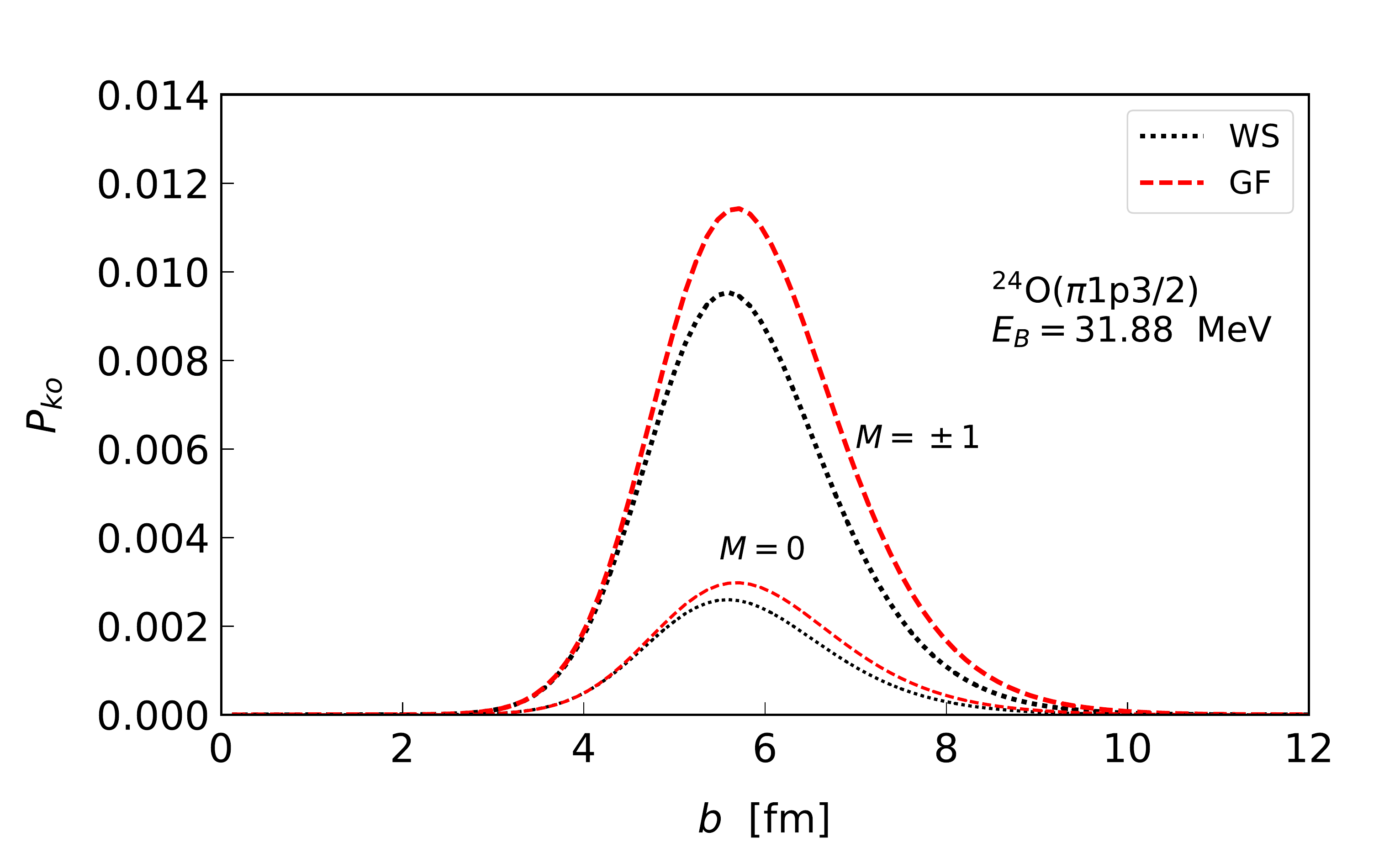}
  \caption{Probability to remove a proton from $^{24}$O, 1p3/2, $E_B = 31.88$ MeV for Wood Saxon (WS, red dashed line) and Green Function (GF, black dotted line) in the $M=0$ (thin lines) and $M=\pm 1$ (thick lines) channels calculated with Eq. (\eqref{eq:Pko}). }
  \label{fig-prob-24O1p32-2}
\end{figure}

We apply Eq. (\ref{eq:bbn}) to obtain the heavy ion proton knockout from $^{24}${\rm O}, 1p3/2, with $E_B = 31.88$ {\rm MeV}. In Fig. \ref{fig-wfOp32-2} we compare the  GF and WS wavefunctions, noticing that while the tails are similar to a Whittaker function (only seen in a logarithmic scale), there are visible differences in their overall shapes. The calculated proton stripping probabilities from Eq. \eqref{eq:Pko} are shown in Fig.  \ref{fig-prob-24O1p32-2}. They are larger for the GF wavefunctions, yielding larger cross sections, as expected by inspecting Table \ref{tab1}.

\begin{figure}[tbp] 
  \centering
  \includegraphics[width=2.7in,keepaspectratio]{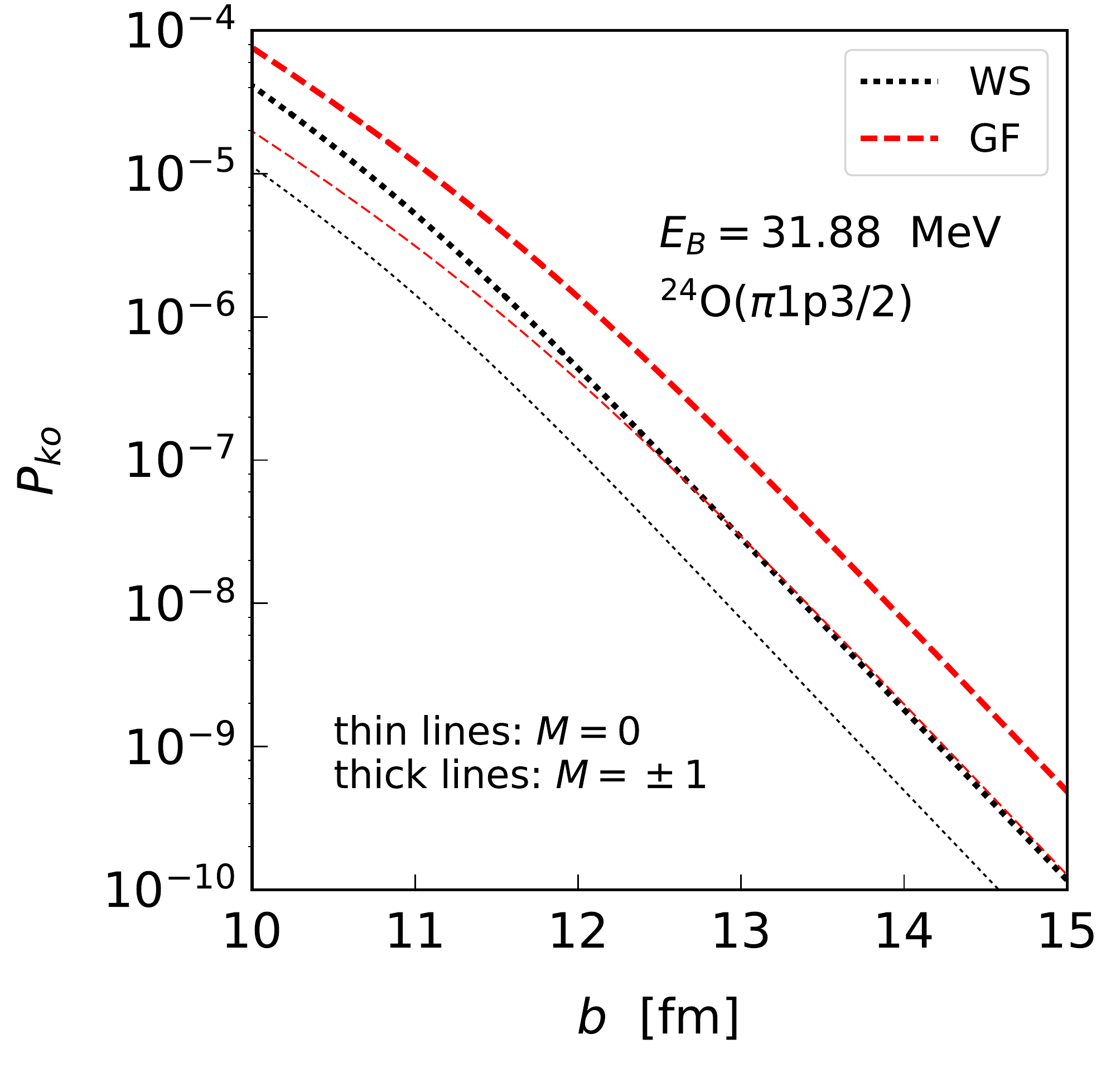}
  \caption{Same as in Fig. \ref{fig-prob-24O1p32-2} but for very large impact parameters $b$, where the integrand in Eq. \eqref{eq:Pko} is dominated by the tail of the s.p. w.f.}
  \label{fig-prob3-24O1p32-2}
\end{figure}

A simple question raises: do the heavy ion knockout cross sections scale with the square of the ANCs? The answer is negative. The respective ANCs scale as $(C_{GF}/C_{WS})^2 \sim 3$, whereas the cross sections scale as $\sigma_{ko}^{GF}/\sigma_{ko}^{WS} \sim 1.25$. This intriguing difference is best understood if the stripping probability is plotted logarithmically for large $b$. This is shown in Fig.  \ref{fig-prob3-24O1p32-2}. While at very large distances, the probability seems as if it scales with a single factor (the ratio between the ANCs), at lower but still large impact parameters they visibly differ from simple scaling. This result is understood by considering the stripping probability in Eq. (\ref{eq:Pko}). Even for large $b$, when the core and the target pass by as much as 10 fm apart, the inner parts of the wavefunction are still probed because the integrand is too small to make substantial contributions to the probability if $b_n \gg 1$, as the $1-|S_n|^2$ goes to rapidly to zero there. We have observed the same behavior for all the cases displayed in Table \ref{tab1}.

\begin{figure}[t] 
  \centering
  \includegraphics[width=2.7in,keepaspectratio]{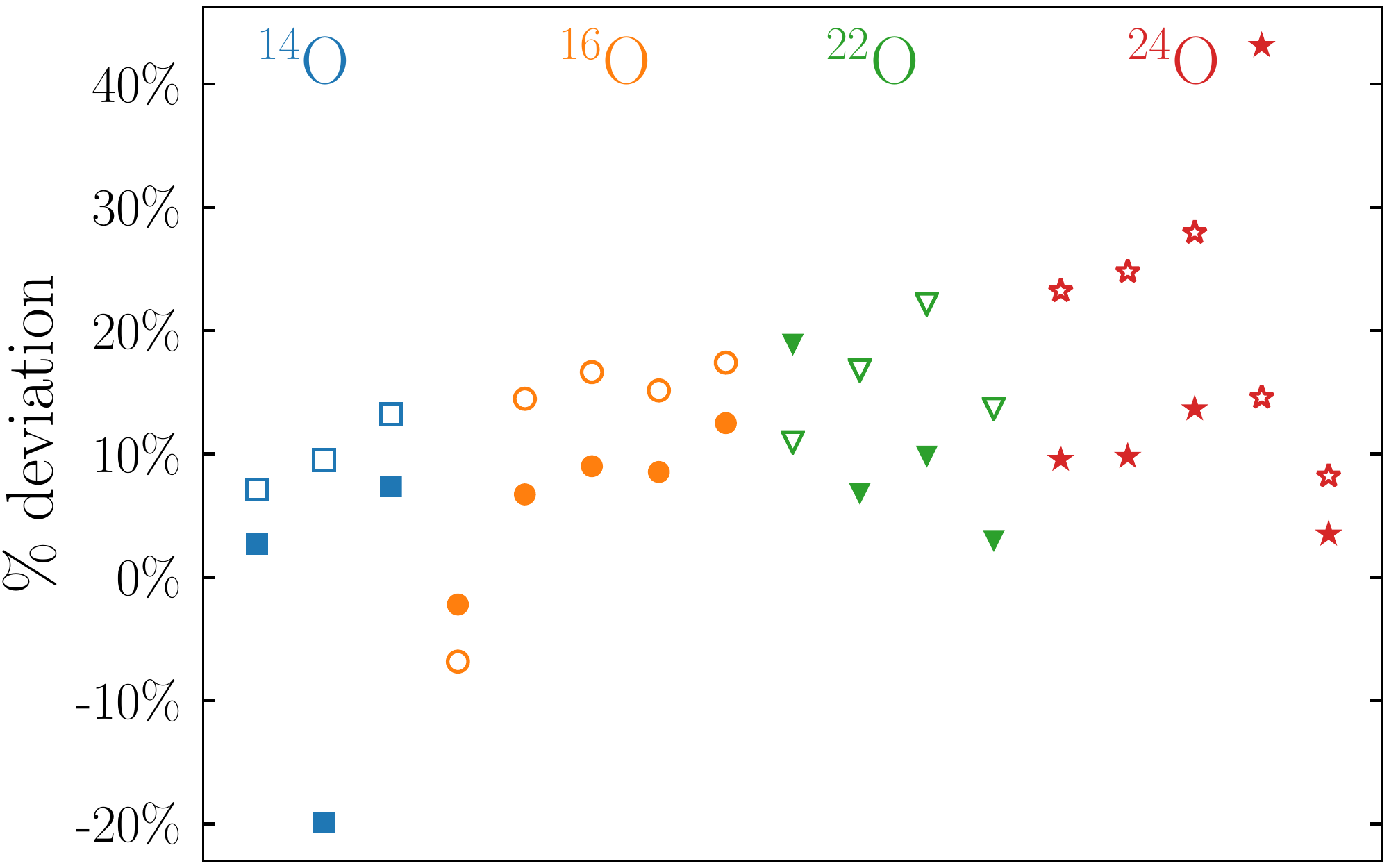}
  \caption{Percent deviation of cross sections using WS wavefunctions and GF overlaps for the 17 reactions in Table \ref{tab1} for QFS (full symbols) and knockout reactions (empty symbols).  Cross sections calculated with  GF overlaps are larger than those with WS wavefunctions, except for two states.}
  \label{fig3}
\end{figure}

The imprints of the details of the many-body overlap functions are  summarized in  Figure \ref{fig3} for the 17 reactions in Table \ref{tab1}. The horizontal scale is a list of the reactions in Table \ref{tab1} from top to bottom of the table. The vertical scale represents $(\sigma_{GF} - \sigma_{WS})/\sigma_{WS}$ in percent for $(p,pN)$ reactions. Except for two cases, the quasi-free cross sections calculated with  GF overlaps are larger than those with WS wavefunctions. The squares (diamonds) [circles] \{stars\} represent these quantities for 350 MeV/nucleon $^{14}$O ($^{16}$O) [$^{22}$O] \{$^{24}$O\} projectiles. It is evident that the results change appreciably with a different form of the internal part of the overlap functions. Fig. \ref{fig3} also demonstrates that variations with respect to the overlap functions are smaller in the (p,pN) case (full symbols). This as due to the capability of this reaction mechanism of better probing the internal part of the nucleus.

{\bf Conclusions.}
In contrast to a commonly considered idea, {\it both} heavy ion knockout reactions and $(p,pN)$ reactions are sensitive to the internal details of the overlap wave function and put strong constraints on the coordinate dependence of the many-body wavefunctions. 

An accurate experimental analysis ideally requires not only the input of an accurately determined overlap function from many-body computations, but also a direct comparison among possible predictions, so that one can assess the extent of the model dependence for the inferred spectroscopic factors. The latter task requires particular attention since a good reproduction of nuclear binding energies and radii is a fundamental constraint but only a fraction of currently available \emph{ab initio} Hamiltonians offer satisfactory saturation properties \cite{10.3389/fphy.2019.00245,10.3389/fphy.2020.00098,10.3389/fphy.2020.00029}.
While this poses a more difficult task for the study of single-particle configurations with heavy-ion
knockout and (p,pN) reactions, it also opens opportunities for a better and more profound understanding of the many-body configurations and their single-particle overlaps. 

In view of the recent advances in experimental facilities and detection techniques, it is suggested that heavy-ion knockout and $(p,pN)$ reactions are analyzed using a consistent many-body model, because they are a formidable tool to extend our knowledge in nuclear spectroscopy only when many-body correlations are considered in the analysis.

{\bf Acknowledgment.} C.B. was supported by the UK-STFC, Grants ST/L005816/1 and  ST/V001108/1. SCGF computations were performed on the DiRAC Data Intensive service at Leicester (funded by the UK BEISvia STFC Capital Grants  ST/K000373/1 and  ST/R002363/1 and STFC DiRAC Operations Grant  ST/R001014/1. C.A.B. was supported by the Helmholtz Research Academy and  the U.S. DOE grant DE- FG02-08ER41533.  A.I. was supported by the Royal Society and Newton Fund with a Newton International Fellowship and the Crafoord Foundation.



%
%


\end{document}